\documentclass[final,3p,times]{elsarticle}

\usepackage[latin1]{inputenc}
\usepackage{multicol}

\usepackage{amssymb}
\usepackage{amsmath}
\usepackage{amsthm}

\usepackage[dvipdfm]{hyperref}
\usepackage{natbib}
\usepackage{graphicx}
\usepackage{pifont}
\usepackage{geometry}
\usepackage{txfonts}

\biboptions{sort&compress}

\journal{Nuclear Physics A}

\begin{document}
\begin{frontmatter}
\title{Universal geometrical scaling for hadronic interactions}
\author{C. Andr\'es}
\author{A. Moscoso}
\author{C. Pajares}
\ead{pajares@fpaxp1.usc.es}
\address{Departamento de F\'isica de Part\'iculas and IGFAE, Universidade de Santiago de Compostela, 15782, Santiago de Compostela, Spain}
\begin{abstract}
It is shown that by defining a suitable saturation momentum $Q_s$, the $p_T$ distributions of pp and AA collisions for any centrality and energy depend only on $\tau=p^2_T/Q_s^2$ for $p_T<Q_s$. The corresponding $\tau$-lines present small differences for $\tau<1$ for different projectiles, targets and centralities. For $\tau>1$, the higher the energy or the larger the size of the participant nuclei, the larger suppression the respective spectra present. The integrated spectrum gives a fraction of the hard multiplicity in the range from 9 \% for pp at 0.9 TeV to 2 \% for Pb-Pb central collisions at 2.76 TeV. 
\end{abstract}
\end{frontmatter}

\section{Introduction}
Recently, it has been shown that in pp collisions the $p_T$ spectra of charged particles exhibit geometrical scaling \cite{ref1,ref2}. Indeed, the $p_T$ spectra in pp collisions in the broad range of energies from 0.9 to 7 TeV scale in a single variable $\tau\equiv p_T^2/Q_s^2$, where the proton saturation momentum $Q_s^p$ is given by
\begin{equation}
\left(Q_s^p\right)^2\equiv Q_0^2\left(\frac{W}{p_T}\right)^\lambda\,,
\end{equation}
where $W=\sqrt{s}\times 10^{-3}$ and $\lambda=0.27$.

In this paper, we study the extension of the geometrical scaling to AA collisions, considering RHIC and LHC energies, different centralities and different nuclei. We show that the geometrical scaling is valid not only for each collision at fixed centrality separately but also for any centrality for $\tau<1$. Furthermore, integrating the hard spectrum for $\tau>1$ we obtain that the hard multiplicity fraction decreases with the size of the participant nuclei as it would be expected from jet quenching. 

\section{The saturation momentum}
In order to check the geometrical scaling in AA collisions for any centrality we define the saturation momentum $Q_s^A$ by
\begin{equation}
\label{par}
\left(Q_s^A\right)^2=\left(Q_s^p\right)^2N_A^{\beta(s)/2}A^{1/6}\left(\frac{A}{N_A}\right)^{1/3}
\end{equation}
where
\begin{equation}
\label{alfa}
\beta(s)=\frac{1}{3}\left(1-\frac{1}{1+\ln\left(\sqrt{s/s_0}+1\right)}\right)\,.
\end{equation}

$N_{A}$ is the number of participant nucleons and $A$ is the number of nucleons.

We observe that at high energy $\beta(s)=1/3$. With this value, for $N_A=A$ we recover the commonly used behaviour of $Q_s^A$ for central collisions $\left(Q_s^A\right)^2=\left(Q_s^p\right)A^{1/3}$.

The parametrization of equation (\ref{par}) is based on the the description of the experimental data on $dN/dy$ of pp and AA at all centralities, rapidities and energies using the framework of percolation of strings \cite{ref3,ref4}. In this approach, the multiplicity distributions of any type of collisions depend on the string density, $\eta_{t}^{A}$, which is a function of the string energy, $s$, the number of participants, $N_{A}$, and the number of nucleons of the nuclei, $A$. This additional dependence on $N_{A}$ and $A$ is crucial to describe the differences observed among the multiplicity spectra of collisions across different nuclei but with the same number of participants. The function $\beta(s)$ shown in formula (\ref{alfa}) takes into account the energy conservation. This effect explains the different exponent of the power-like growing of the energy dependence of the pp and AA central multiplicities. Usually, in strings or colour flux tube models, the number of strings formed in AA collisions is proportional to ${N_{A}}^{4/3}N_{s}$, being $N_{A}^{4/3}$ the number of collisions and $N_{s}$ the number of strings formed in a pp collision at the same energy. However, the formation of one string requires a minimum of energy, energy of around 0.5-0.6 GeV. As the total available energy is $A\sqrt{s}$, at low and intermediate energies, there is not enough energy to be shared by such a large number of strings. We expect a suppression of the power 4/3 to $1+\beta(s)$, with $\beta(s)\rightarrow 1/3$ at high energy. Moreover, as the energy of each string is related to the longitudinal phase space that grows like $\ln s$, the approach to the asymptotic limit 1/3 should be logarithmic as the parametrization of formula (\ref{alfa}). 

In addition, the string percolation approach is related to the glasma of the colour glass condensate \cite{ref19} and there is a correspondence between the number of clusters of strings (effective number of colour sources), $\eta^{1/2}R_{A}^{2}$, and the number of colour flux tubes of the glasma, $(Q_{s}^{A})^{2}R_{A}^{2}$. In this way, the dependence of the string density $\eta_{t}$ on $s$, $A$ and $N_A$ is translated into $(Q_{s}^{A})^{2}$, yielding in equation (\ref{par}).

The dependence of $(Q_{s}^{A})^{2}$ on $s$, $A$ and $N_A$ establishes the relative normalization of the transverse momentum distributions of different collisions (the integrated distributions are proportional to $(Q_{s}^{A})^{2}R_{A}^{2}$).

The values of the parameters $Q_{0} = 1 GeV$ and $\lambda = 0.27$ are taken from the reference \cite{ref1}. A slightly worse scaling has been obtained with $\lambda = 0.30$. The value of $\sqrt{s_{0}} = 245 GeV$ was obtained in references \cite{ref3,ref4} and indicates the energy scale of the energy conservation effect.

The different energy dependence of the saturation momentum for pp and AA collisions is also obtained in the CGC framework. Indeed, it has been shown that the DGLAP and the BK evolution equations, with a detailed inclusion in the nuclear geometry, yield a different evolution speed \cite{ref20, ref21} able to describe the pp and AA multiplicity data. It has been also pointed out that in addition to the $(Q_{s}^{A})^{2}R_{A}^{2}$ dependence, the multiplicity distribution has a factor describing the fragmentation of gluons into hadrons, $N_{h}^{g}(Q_{s}^{2})$. According to the $e^{+}e^{-}$ data on jet production, this fragmentation function is constant for $Q^{2} < 1 (GeV)^2$ but grows for $Q^{2} > 1 (GeV)^2$. In pp and AA collisions we have $Q_{s}^{p} < 1 GeV$ and $Q_{s}^{A} \geq 1 GeV$ respectively, therefore in the case of AA collisions there is an additional energy dependence which originates a different increasing of pp and AA multiplicities \cite{ref22}. The equation \ref{par} can be regarded as a parametrization 
of these effects.

Finally, in equation (\ref{par}), $(Q_{s}^{A})^2$ does not depend on rapidity; we will nevertheless use it only in the central pseudorapidity region. In most of the string models, strings stretching among sea quarks and antiquarks are expanded only in the central pseudorapidity range. Outside this pseudorapidity range, there are only strings involving valence quarks or valence diquarks. This fact changes the dependence of the number of strings on $N_{A}$. For central pseudorapidity, $N_{s} \sim N_{A}^{4/3}$ and outside this range $N_{s} \sim N_{A}$. This change gives rise to a smoother dependence of $(Q_{s}^{A})^{2}$ on $N_{A}$ outside the central pseudorapidity region.

\section{Comparison with experimental data (RHIC, LHC)}
\begin{figure*}
\begin{center}
\includegraphics[scale=0.3]{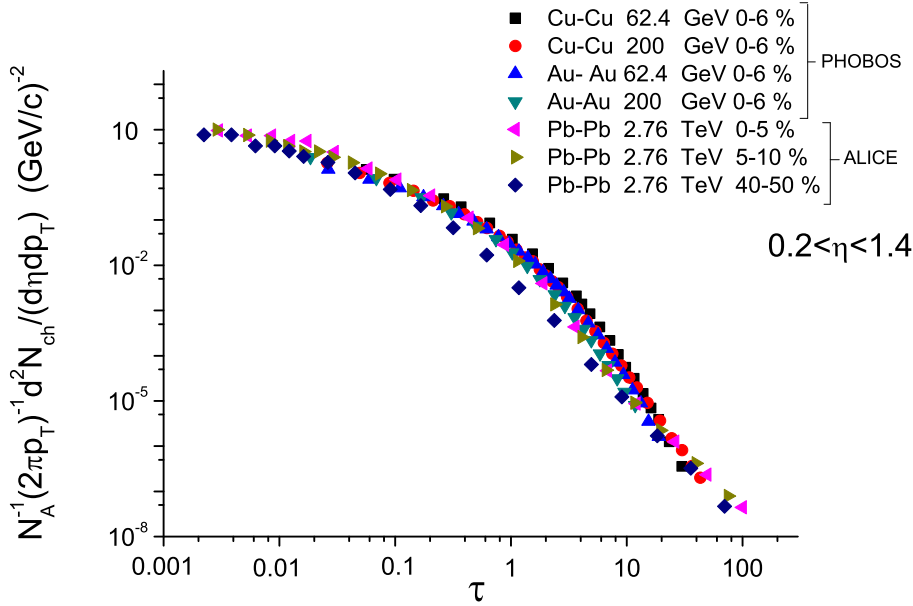}
\caption{(Colour online.) Charged particle multiplicity per participant at pseudorapidity $0.2<\eta<1.4$ for Au-Au and Cu-Cu central collisions at two RHIC energies 62.4 and 200 GeV \cite{ref8,ref9} and for Pb-Pb collisions at 2.76 TeV \cite{ref10} plotted as a function of $\tau$ for $\lambda=0.27$.}
\label{fig1}
\end{center}
\end{figure*}

In Fig. \ref{fig1} we show the data $(N_A2\pi p_T)^{-1}d^2N_{ch}/(dp_T d\eta)$ for Cu-Cu 0-6 \% central at 62.4 and 200 GeV and Au-Au 0-6 \% central at 62.4 and 200 GeV \cite{ref8,ref9} together with Pb-Pb 40-50 \%, 5-10 \% and 0-5 \% at 2.76 TeV \cite{ref10} as a function of $\tau$, in the pseudorapidity range $0.2<\eta<1.4$. The number of participants used is the mean value corresponding to the given centrality. Experimental data taken from ALICE \cite{ref10} correspond to a pseudorapidity range (including the $\eta=0$ region) in which $dn/d\eta$ is smaller than in the pseudorapidity range considered here. Because of this, a 15 \% correction was applied to the normalization \cite{ref4}.

It can be seen that Pb-Pb 0-5 \% and 5-10 \% data points fall into the same line for $\tau<1$. Also the Cu-Cu and the Au-Au data points at both energies are approximately in the same curve for $\tau<1$. Only the Pb-Pb 40-50 \% points present some departure around 1. For $\tau>1$ it is shown a suppression for the heavy nuclei.

\begin{figure*}
\begin{center}
\includegraphics[scale=0.3]{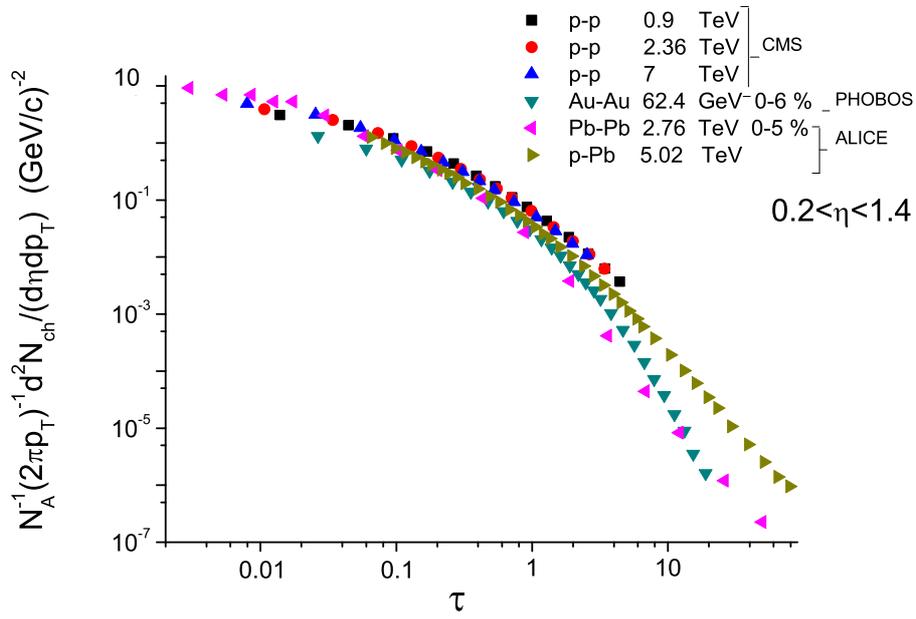}
\caption{(Colour online.) Charged particle multiplicity per participant at pseudorapidity $0.2<\eta<1.4$ for pp collisions \cite{ref5, ref6, ref7}, Au-Au 0-6 \% central collisions at 62.4 GeV \cite{ref8}, Pb-Pb 0-5 \% collisions at 2.76 TeV \cite{ref10} and p-Pb data at 5.02 TeV \cite{ref11} versus $\tau$ for $\lambda=0.27$.}
\label{fig2}
\end{center}
\end{figure*}

In order to see the differences between the different sizes of projectile and target, Fig. \ref{fig2} shows the data for pp collisions at 0.9, 2.36 and 7 TeV \cite{ref5, ref6, ref7} together with the recent p-Pb data at 5.02 TeV \cite{ref11}, the Au-Au 0-6 \% central data at 62.4 GeV \cite{ref8} and the Pb-Pb 0-5 \% data at 2.76 TeV \cite{ref10} in the pseudorapidity range $0.2<\eta<1.4$. For the p-Pb data we use $<N_{part}>=7.9$ as quoted in reference \cite{ref12}.

\begin{figure*}
\begin{center}
\includegraphics[scale=0.3]{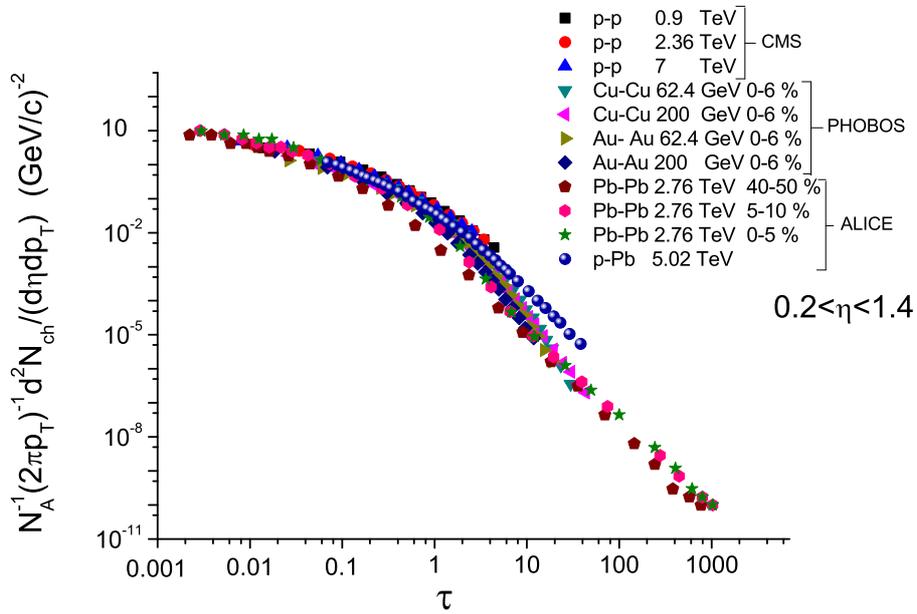}
\caption{(Colour online.) Charged particle multiplicity per participant in the pseudorapidity range $0.2 < \eta < 1.4$ for all the heavy ion collisions considered plotted as a function of $\tau$ for $\lambda = 0.27$.
}
\label{fig3}
\end{center}
\end{figure*}

We observe that the pp data at different energies fall into the same line, satisfying geometrical scaling, as it was shown in reference \cite{ref1}. Furthermore, we observe that the p-Pb, Pb-Pb 0-5 \% and Au-Au 0-6 \% central data at low $\tau$ are very close to the pp data of different energies. As $\tau$ becomes larger the difference between these sets of data increases. For $\tau>1$ the suppression is larger for Pb-Pb 0-5 \% central than for p-Pb, which still is more suppressed than for pp data.

The scaling for $Q^{2} < Q_{s}^{2}$ was predicted by CGC and by phenomenological saturation models \cite{ref23, ref24, ref25, ref26}. The HERA data on deep inelastic scattering at low x show scaling even for very high $Q^{2}$, $Q^{2} < 400 GeV^{2}$ \cite{ref27}. Indeed, the solution of the BFKL evolution equation shows that the scaling can be extended to intermediate $Q^{2}$, $1 \lesssim ln (Q^{2}/Q_{s}^{2}) \ll ln (Q_{s}^{2}/\Lambda_{QCD}^{2})$ \cite{ref28}. The comparison with data shows that pp data lie in the same curve at different energies even for $\tau > 1$. On the contrary, this is not true for AA collisions. Notice that in the pp case we do not expect jet quenching because we do not have a high density medium, in opposition to AA collisions (at high LHC energies and high multiplicity events, jet quenching has been predicted \cite{ref29, ref30} but the weight of these events compared to minimum bias is negligible).

In Fig. \ref{fig3}, all previous data are plotted to show that for $\tau < 1$ there are not many differences among them, which does not occur for the $\tau >1 $ region, where suppression is larger for larger sizes of nuclei.

\begin{figure*}
\begin{center}
\includegraphics[scale=0.3]{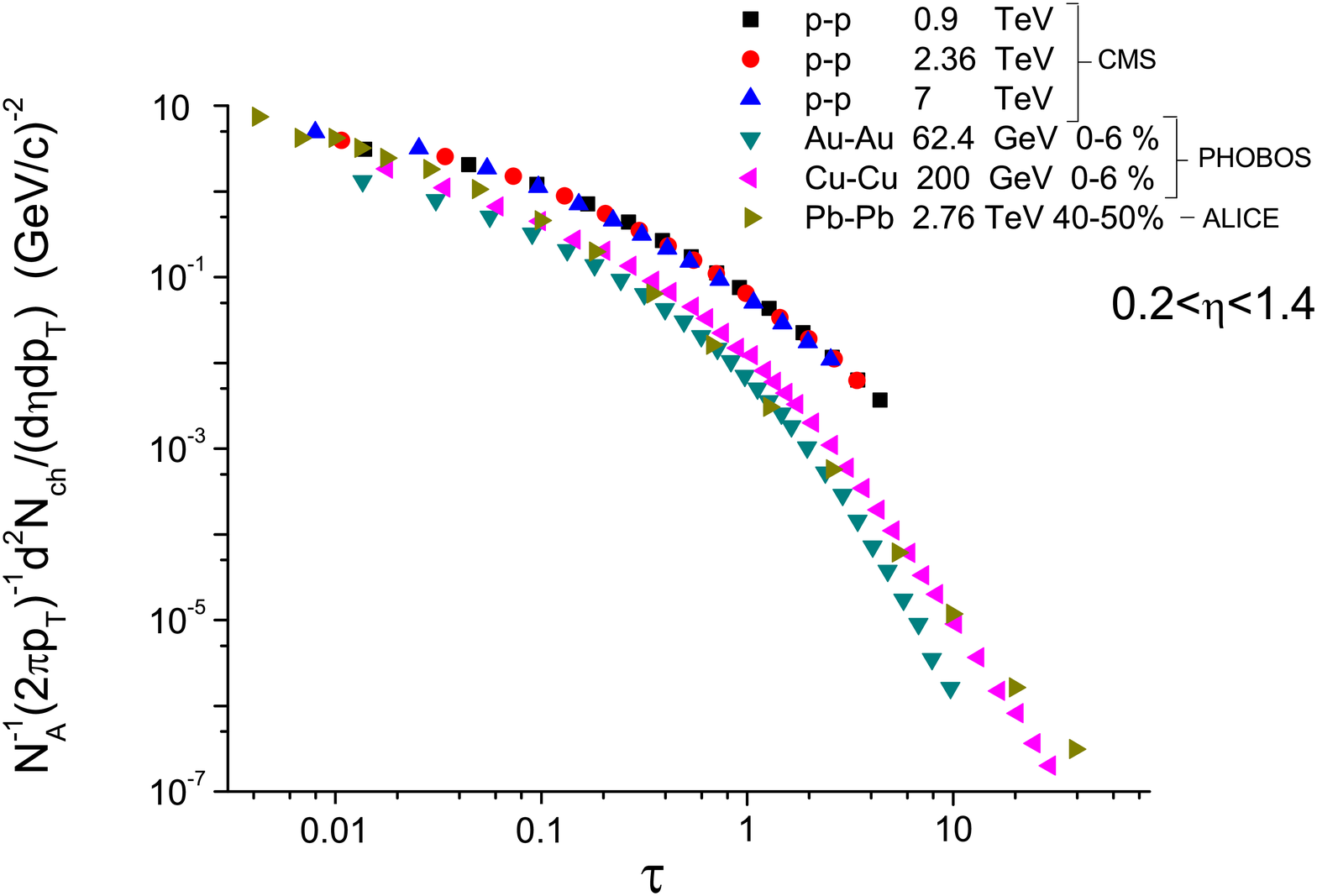}
\caption{(Colour online.) Charged particle multiplicity per participant in the pseudorapidity range $0.2 < \eta < 1.4$ for pp collisions \cite{ref5, ref6, ref7}, Au-Au 0-6 \% central collisions at 62.4 GeV \cite{ref8}, Cu-Cu 0-6\% central collisions at 200 GeV \cite{ref9} and Pb-Pb 40-50 \% collisions at 2.76 TeV \cite{ref10} versus $\tau$ for $\lambda=0.27$ using $\beta(s) = 1/3$.}
\label{fig5}
\end{center}
\end{figure*}

We emphasize that the parametrization (\ref{par}) allows us to deal with any type of collisions at any energy. For that purpose, the introduction of the $\beta(s)$ function is required, which implies a different power for different number of participants. In 
reference \cite{ref2} the geometrical scaling was shown for pp collisions at LHC and AA collisions at RHIC using an energy independent exponent $\lambda$, but this exponent was different for pp and AA collisions. If we use an energy independent exponent for both, pp and AA collisions, we spoil our scaling as it can be seen in In Fig. \ref{fig5} where we plot the pp and several AA transverse momentum distributions using $\beta(s) = 1/3$.

\begin{figure*}
\begin{center}
\includegraphics[scale=0.3]{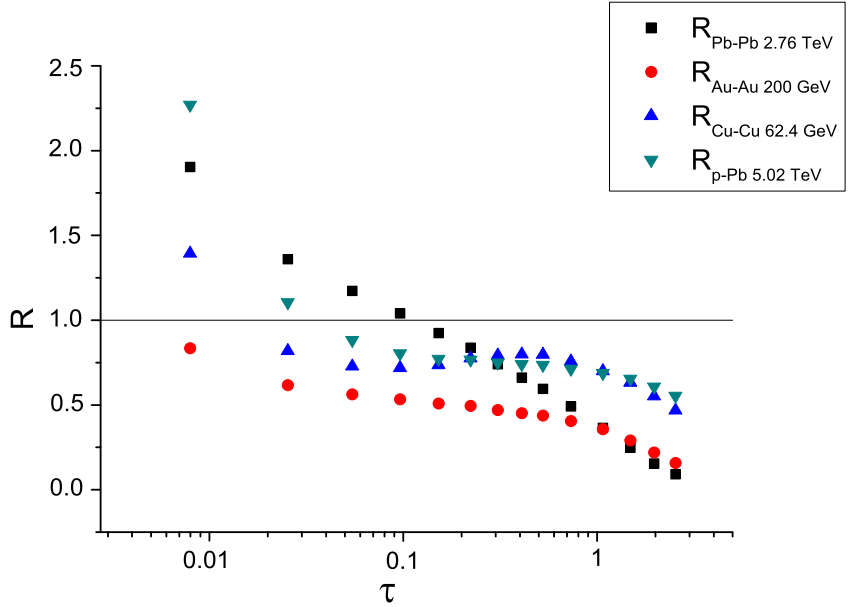}
\caption{(Colour online.) Ratio of Pb-Pb 0-5 \% at 2.76 TeV \cite{ref10}, Au-Au 0-6 \% central at 200 GeV \cite{ref8}, Cu-Cu 0-6\% central at 62.4 GeV \cite{ref9}, pPb at 5.02 TeV \cite{ref11} over pp at 7 TeV \cite{ref9} in terms of $\tau$.
}
\label{fig4}
\end{center}
\end{figure*}

In order to see the quality of this extended scaling we show in Fig. \ref{fig4} the ratio of Pb-Pb 0-5 \% at 2.76 TeV, Au-Au 0-6 \% at 200 GeV, Cu-Cu 0-6\% at 62.4 GeV, pPb at 5.02 TeV over pp at 7 TeV as a function of $\tau$. Even for such different projectiles and targets and also in the broad range of energies considered, we observe an approximate scaling at low $\tau$. Notice that for $\tau<1$ the data extend over three orders of magnitude. Ratios vary between 0.5 and 1 in the range $0.2<\tau<1$. Notice that for $p_{T} < \Lambda_{QCD}$, or equivalently for $\tau < 0.1-0.2$, there is no reason to expect geometrical scaling. The violation of the scaling is clear for $\tau>1$, showing a higher suppression for heavier nuclei. The hierarchy of the scaling violations for $\tau>1$ agrees with the expected suppression due to jet quenching.

Based on this approximate scaling, we can compute the multiplicity of soft and hard particles per participant, defining soft particles as those with $p_T<Q_s$ and hard particles as those with $p_T>Q_s$. In fact,
\begin{equation}
\frac{1}{N_A}\frac{dN_{ch}^{soft}}{d\eta}=\frac{1}{N_A}\int_0^{Q_s^2}\! \mathrm{d}p_T^2 \, \frac{dN^{2}_{ch}}{d\eta dp_T^2}=\frac{1}{N_A}\int_0^{Q_s^2}\! \mathrm{d}p_T^2 \, \frac{1}{Q_0^2} F(\tau)
\end{equation}
and with the change of variables 
\begin{equation}
\frac{dp_T^2}{Q_0^2}=\frac{2}{2+\lambda}\left(\frac{W}{Q_0}\right)^{\frac{2\lambda}{2+\lambda}}\tau^{-\frac{\lambda}{2+\lambda}}N_A^{\beta(s)/2}A^{1/6}\left(\frac{A}{N_A}\right)^{1/3} d\tau,
\end{equation}
the fraction of soft and hard multiplicities over the total multiplicity results 
\begin{equation}
\label{6}
R_s\equiv \frac{dN_{ch}^{soft}/d\eta}{dN_{ch}^{tot}/d\eta}=\frac{\int_0^1\! \mathrm{d}\tau \, \tau^{-\frac{\lambda}{2+\lambda}} F(\tau)}{\int_0^\infty\! \mathrm{d}\tau \, \tau^{-\frac{\lambda}{2+\lambda}}F(\tau)}
\,,\qquad
R_h\equiv \frac{dN_{ch}^{hard}/d\eta}{dN_{ch}^{tot}/d\eta}=\frac{\int_1^\infty\! \mathrm{d}\tau \, \tau^{-\frac{\lambda}{2+\lambda}} F(\tau)}{\int_0^\infty\! \mathrm{d}\tau \, \tau^{-\frac{\lambda}{2+\lambda}}F(\tau)}
\end{equation}

In Table \ref{tab1} we present the results for the fractions of soft and hard multiplicities for the centralities, energies and collisions considered. In order to do the integration in (\ref{6}), we perform a fit to each $p_{T}$-distribution separately which is afterwards integrated.

We observe that the hard fraction decreases slowly with energy and also decreases with the size of the participant nuclei. In the case of pp collisions the dependence of the energy is very weak, varying from 9 \% to 8 \% in the broad range 0.9-7 TeV. However, we observe larger differences between Au-Au 0-6 \% central at 62.4 GeV, whose hard fraction is 7 \%, and Pb-Pb 0-5 \% at 2.76 TeV, which is 2 \%. For Pb-Pb at 2.76 TeV the hard fraction is the same for peripheral (40-50 \%) than for central (0-5 \%).

\begin{table}[h]
\begin{center}
\begin{tabular}{c c c c c}
 & Energy (TeV)  & Centrality & $R_s$  & $R_h$ \\
\hline
p-p & 0.9 & minb  & 0.91 & 0.09\\
p-p & 2.36 & minb & 0.90 & 0.10\\
p-p & 7 & minb & 0.92 & 0.08\\
p-Pb & 5.02 & minb & 0.93 & 0.07\\
Cu-Cu & 0.0624 & 0-6 \% & 0.93 & 0.07\\
Cu-Cu & 0.2 & 0-6 \% & 0.94 & 0.06\\
Au-Au & 0.0624 & 0-6 \% & 0.93 & 0.07\\
Au-Au & 0.2 & 0-6 \% & 0.96 & 0.04\\
Pb-Pb & 2.76 & 40-50 \% & 0.98 & 0.02\\
Pb-Pb & 2.76 & 0-5 \% & 0.98 & 0.02\\
\end{tabular}
\end{center}
\caption{\label{tab1}Values of the fractions of soft and hard multiplicities for different nuclei, centralities and energies.}
\end{table}

\section{Conclusions}

Usually it is believed that as the energy or the size of the participant nuclei increases, the weight of the hard collisions increases and, correspondingly, the weight of the hard multiplicity. We observe the opposite with our definition of hard and soft collisions. This result was also pointed out by the ALICE collaboration data of the sphericity as a function of the energy and the change multiplicity in pp collisions at $0.9$, $2.36$ and $7$ TeV \cite{ref13,ref14}. The sphericity measures the jet activity, in such a way that one event with dijet back to back implies sphericity zero. On the contrary, an event where all the produced particles are distributed isotropically in phase space implies sphericity one. Most of the Monte-Carlo codes (Pythia 8, Perugia-0, Phojet, Atlas-CSC) predict a decreasing of the sphericity with energy and charged multiplicity whereas the data show the opposite trend. In the color glass condensate model \cite{ref15, ref16} the more the saturation momentum grows with the energy and the number of participants, the smaller the room for hard collisions. This also happens in the framework of percolation strings \cite{ref17,ref18} where the area covered by strings increases with the size and energy of the nuclei leaving less room for hard scatterings.

Let us mention that the scaling found is naturally incorporated in the colour glass condensate \cite{ref15, ref16} where the $p_{T}$ spectrum depends only on $p^{2}_{T}/Q^{2}_{S}$. In the model of percolation of strings \cite{ref17, ref18} a scaling law for the $p_{T}$ dependence is also obtained. Now the role of $Q^{2}_{S}$ is played by $\sqrt{\eta^{t}}$ where $\eta^{t}$ is the string density \cite{ref19}. Indeed the parametrization (\ref{par}) has its origin in the parametrization of $\sqrt{\eta^{t}}$ \cite{ref3}.

In conclusion, we have shown that the geometrical scaling previously seen in pp collisions can be extended to AA collisions, being satisfied also for different centralities. For different projectiles and targets an approximate scaling is also satisfied although in this case some differences occur. The departure of scaling for $\tau>1$ follows the expected hierarchy in jet quenching, namely the suppression is larger for larger participant nuclei.

\section*{Acknowledgements}

\hspace{1.5em}We thank L. McLerran, N. Armesto and C. Salgado for useful discussions. This work was supported by the Ministerio de Econom\'ia y Competitividad of Spain under the project FPA 2011-22776, by the Spanish Consolider CPAN project and by Xunta de Galicia.

\end{document}